\documentclass[twocolumn,showpacs,preprintnumbers,amsmath,amssymb]{revtex4}

\usepackage{epsfig}
\usepackage{graphicx}
\usepackage{dcolumn}
\usepackage{bm}

\begin{document}

\title{The influence of spin-dependent phases of tunneling electrons on
the conductance of a point  ferromagnet/isolator/d-wave
superconductor contact}
\author{B.P. Vodopyanov}
\affiliation{Kazan Physical-Technical Institute, Russian Academy
of Sciences, 10/7 Sibirsky Tract, 420029 Kazan, Russia}
\date{\today}

\begin{abstract}
The influence of phase shifts of electron waves passing through
and reflected by the potential barrier on the Andreev reflection
in a ferromagnet/isolator/d-wave superconductor (FIS) contact is
studied.  It is found that in a superconductor the surface
spin-dependent Andreev bound states inside the superconducting gap
are formed as a result of the interference of electron-like and
hole-like quasiparticles due to  repeated Andreev reflections. The
peak in the conductance of the FIS contact at the zero potential
for the (110)-oriented superconductor disappears rapidly as the
polarization of a ferromagnet increases, whereas for the
(100)-oriented superconductor it appears.  The physical reason for
this behavior of conductance is discussed.
\end{abstract}

\pacs{74.50.+r, 74.80.-g, 74.80.Dm, 74.80.Fp, 75.30.Et}

\maketitle

\section{\qquad Introduction}
The presence in a ferromagnetic metal (F) of electron spin
subbands with different values of the Fermi momenta:
$p_{\,\uparrow}$ - for a subzone with the projection $\,\alpha$
($\,\alpha = \uparrow, \downarrow $) of the electron spin up onto
the direction of the magnetic moment of a ferromagnetic and
$p_{\,\downarrow}$ - for a subband with the projection of the
electron spin down results in the oscillating character of the
spatial dependence of the anomalous Green function (GF) in a
ferromagnet  in various hybrid structures containing the interface
ferromagnet/superconductor (F/S) with a singlet order parameter
\cite{BuzdinJTPL, Demler, Kupriyanov, BuzdinRMP}. Such
manifestation of the proximity effect is the basis of the creation
of the $\pi$-Josephson junction \cite{Ryazanov1}, various
spin-valve schemes \cite{BeasleyAPL, Johnson, Tagirov, BuzdinEl,
Nazarov}, being the main elements of promising superconducting
electronics \cite{BeasleyIEEE, Ustinov, Ioffe}.

The presence of spin subbands in a ferromagnetic metal results in
the suppression of the Andreev reflection \cite{Beenakker} in
point F/S contacts; the dependence of the transmission
coefficients $D_{\,\alpha}$ and reflections coefficients $
R_{\,\alpha}\,=\,1-D_{\,\alpha} $ on the projection --$\,\alpha$
of the spin of the quasi particle passing throught the interface
F/S. These properties are used to determine the spin polarization
of ferromagnetic materials \cite{ Buhrman1, Soulen, Perez,
Krivoruchko, Buhrman2}, to study the order parameter symmetry of
high-temperature superconductors \cite{Vasko, Dong, Chen, Luo} and
to control the spin polarized currents \cite{Jedema, Urech,
Appelbaum}.

Recently attention was paid to another property of hybrid  F/S
structures: the influence of spin-dependent phase shifts of
amplitudes of electron states on the Fermi surface $d_{\,\alpha} $
passing through the interface and   $r_{\,\alpha}$
($d_{\,\alpha}\,=\,\sqrt{D_{\,\alpha}}\,\,exp\,(i\,\theta_{\,\alpha}^{\,d}));\,r_{\,\alpha}\,=\,
\sqrt{R_{\,\alpha}}\,\,exp\,(i\,\theta_{\,\alpha}^{\,r})$)
reflected from it on the thermodynamic \cite{Sauls1988} and
transport \cite{Fogelstrom, Barash, Sauls2004} characteristics of
hybrid structures with a spin-active  interface. It was found that
the difference of  spin-dependent phase shifts due to the
difference of the potential barriers for electrons with different
spin projections $\,\alpha$ results in the appearance of a $\pi$
state in the  S/FI/S junction (FI is ferromagnetic isolator)
without accounting for the proximity effect. The presence of
spin-dependent phase shifts also leads to the appearance of
Andreev bound states in ballistic contacts N/FI/S и N/F/S (N is
normal metal) \cite{Sauls2004} resulting in the peak in the
dependence of conductance on the applied potential  $V$. In
\cite{Cottet}, \cite{Vod} it was shown that spin-dependent phase
shifts induce subgap resonances n the conductance FNS
\cite{Cottet} and  FIS (I is isolator) \cite{Vod} in ballistic
contacts as well.

This strong influence of apin-dependent phase shifts on the
transport properties of hybrid structures with ferromagnetic
elements allows one to suppose that they may be used successfully
in experiments on Andreev spectroscopy of ferromagnets,
superconductors and in various applications in the field of
nanospintronics.

This paper is concerned with a theoretical study of the influence
of  spin-dependent phase shifts on the Andreev reflection and
conductance of a point F/I/d-wave superconductor contact.

Superconductors with the d-wave symmetry (superconductors with the
$ d_{x^2 - y^2}$ symmetry of the order parameter are considered)
have an internal, momentum-dependent phase, which strongly
influences the transport properties of contacts between them and
other materials. In paper \cite{Hu} it was shown that when the
angle $\gamma $ between the  а - axis of a superconducting crystal
and the normal to the surface of the high-ohm  interface is $ \pi
/4 $, then a bound state is formed in the Fermi level close to the
high-ohm interface . This zero-energy bound state resulting from
the repeated Andreev reflections \cite{Harlingen},
\cite{Kashiwaya2} causes a sharp peak in the dependence of
conductance of the N/I/d-wave superconductor on the applied
potential at the zero potential  \cite{Kashiwaya}. This peak at
the zero potential is also manifested in the dependence of
conductance on the applied potential at spin-polarized current as
well \cite{Zutic}, \cite{Vod2005}.

The main result of this paper is that  spin-dependent phase shifts
considerably reconstruct the spectrum of conductance of the
F/I/d-wave superconductor contact for the (100) and (110)
orientations of the d-wave superconductor. This takes place
because due to the interference of a part of trajectories of
electron-like and hole-like quasi particles reflected by the pair
potential and interface Fermi bound states on the Fermi level are
formed, whereas due to the interference of the other part of
trajectories bound states are formed in the vicinity of the edges
of the energy gap. This work illustrates that the study of the
influence of spin-dependent phase shifts on the conductance of the
point F/I/d-wave superconductor contact can provide an interesting
insight in spin-dependent transport.

The theoretical possibility to study the influence of the
spin-dependent phase shifts of the amplitudes of electron waves on
the Fermi surface passing through  and reflected from the
interface on the $I\,-\,V$ characteristics of superconducting weak
links  with ferromagnetic elements appearing after the boundary
conditions (BCs) for the quasiclassical GF wasobtained. In paper
\cite{Sauls1}, BCs for the quasiclassical GF for two metals in
contact via a magnetically active interface in terms of an
interface scattering matrix were derived. In paper \cite{Fogel},
BCs for the retarded and advanced quasiclassical GFs were obtained
in terms of Riccati amplitudes \cite{Eschrig}, \cite{Shelankov}.
In paper \cite{Sauls2004}, BCs in terms  of Riccati amplitudes
were obtained for the nonequilibrium quasiclassical GF.   In
papers \cite{Vod2003} quasiclassical equations of
superconductivity for metals with a spin-split conduction band
were derived and BCs for the temperature quasiclassical GF for the
F/S interface were obtained. The model interface was the same as
in  \cite{Sauls1}, \cite{Zaitsev}.

In this paper, calculation are carried out by a quasiclassical GF
with the usage of the relevant BCs obtained in \cite{Vod2003}.

\section{ Finding differential conductance of a point FIS contact}
\subsection{\label{sec:level1} The general expression for differential
conductance of a point contact through quasiclassical GF} In
hybrid F/S structures Andreev reflection is modified. The
reflected hole has some parameters (for example, the velocity
modulus and phase shift) different from those of the incident
electron because it moves in a subband with the opposite spin.Such
spin-discriminating processes due to the exchange field in
aferromagnet lead to the formation of Andreev bound states inside
the gap \,\cite{Barash},\,\cite{Fogel}.

The enegy of Andreev bound states depends on the spin index
\cite{Barash},\,\cite{Fogel}. As a result, the spectral density of
conductance $ G_{FIS} $ of the FIS contact at zero voltage is no
longer  a symmetrical function of energy $\varepsilon$. The
condition of the time reversal invariance has the form
$G_{FIS}(\varepsilon,\,\alpha) $\,=\,$
G_{FIS}(-\,\varepsilon,\,-\,\alpha) $.  The generalization of the
conductance $ G_{FIS}(V)$ \,\, \cite{Tinkham} for this case
\cite{Vod} results in the following expression for  $ G_{FIS}(V)$:
\begin{eqnarray}
G_{FIS}(V)=\frac{e^{\,2} A}{32\pi \,T}\sum_\alpha\, \rm{Tr} \left[
\int\frac{d{\bf p}_\|}{\,(2\,\pi)^{\,2}}
\int\limits_{-\infty}^{\infty} d\varepsilon\,\times \right. \notag
\\
\frac{1}{\coth^2(\frac{\varepsilon-eV\,\,
\hat{\tau}_z}{2\,T})}\,\, [1-\hat{g}_{\,s}^{\,A}\, \hat{\tau}_z
\,\hat{g}_{\,s}^{\,R} \,\hat{\tau}_{\,z}
-\hat{g}_{\,a}^{\,A}\,\hat{\tau}_z
\,\hat{g}_{\,a}^{\,R}\,\hat{\tau}_{\,z} \notag
\\
\left. +\hat{\Upsilon}_{\,s}^{\,A}\hat{\tau}_z
\hat{\Upsilon}_{\,s}^{\,R}\,\hat{\tau}_{\,z}-\hat{\Upsilon}_{\,a}^{\,A}
\hat{\tau}_z \hat{\Upsilon}_{\,a}^{\,R} \hat{\tau}_{\,z}]\right].
\label{eq:1}
\end{eqnarray}
In Eq. (\ref{eq:1}), \,$A$\, is the contact area; $\hat{\tau}_z$
is the Pauli matrix; \, $p_\|$ is the momentum in the contact
plane;\, ($\hat{g}_{\,s}$, $\hat{\Upsilon}_{\,s}$) and
\,($\hat{g}_{\,a} $, $\hat{\Upsilon}_{\,a}$) are quasiclassical
retarded (R) and advanced (A) GFs symmetric (s) and antisymmetric
(a) \,\cite{Vod} with respect to the projection of the momentum
\,$ {\bf\hat{p}}$ \, on the Fermi surface on the axis $x$,
Perpendicular to the contact plane, composed according to the rule
$\hat{T}_{\,s(a)}\,=\,1/2\,[\hat{T}( p_{\,x})\,\pm \,\hat{T}(-\,
p_{\,x})]$. Calculations in Eq. (\ref{eq:1}) are to be carried out
on the boundary of any contacting metal.

\subsection{\label{sec:level2} Finding quasiclassical GF}
Let us assume that the barrier with the width $d$ is located in
the region $ -\,d/2\,<x<\,d/2$, the superconductor occupies the
region $x>d/2$, and the ferromagnet occupies the region
$x<-\,d/2$. To find GFs, for each metal one has to solve
quasiclassical equations of superconductivity for metals with a
spin-split conductivity band simultaneously with their BCs derived
in paper \cite{Vod}:
\begin{multline}\label{eq:2}
{\rm{sign}}(\hat{p}_{\,x})\frac{\partial}{\partial{\,x}}\,\hat{g}+\frac{1}{2}\,
{\bf
v_\|}\frac{\partial}{\partial{\bf\rho}}(\hat{v}_{\,x}^{-1}\hat{g}+\hat{g}\,\hat{v}_{\,x}^{-1})
+[\hat{K},\,\hat{g}]_- = 0, \\
{\rm
{sign}}(\hat{p}_{x})\frac{\partial}{\partial{x}}\hat{\Upsilon}+\frac{1}{2}
{\bf
v_\|}\frac{\partial}{\partial{\bf\rho}}(\hat{v}_{x}^{-1}\hat{\Upsilon}-
\hat{\Upsilon}\hat{v}_{x}^{-1})+[\hat{K},\hat{\Upsilon}]_+ = 0,\\
\hat{K}=\,-\,i\hat{v}_{\,x}^{-\frac{1}{2}}(i\varepsilon_n
\hat{\tau}_z+
\hat{\Delta}-\hat{\Sigma})\hat{v}_{\,x}^{-\frac{1}{2}} -
i(\hat{p}_{\,x}-\hat{\tau}_x\hat{p}_{\,x}\hat{\tau}_x)/2, \\
\hat{\Delta}\,\equiv \,\hat{\Delta}(x,\,{\bf p}),\qquad
\shoveleft{[a,\,b]_\pm = ab \pm ba.} \qquad
\end{multline}
In this section, $\varepsilon_n = (2n + 1)\pi T $  is the
Matsubara frequency; $\hat{\Sigma}$ is the self-energy part;
$\hat{g}$  are matrix temperature GFs:
$$ \hat{g}=
  \begin{pmatrix}
    g_{\,\alpha,\,\alpha} & f_{\,\alpha,\,-\alpha} \\
    f_{-\,\alpha,\,\alpha}^{\,+}&\,-\, g_{-\,\alpha,\,-\alpha}
  \end{pmatrix}, \,\,
 \hat{g}=\left\{\begin{split}\hat{g}_>\qquad  \hat{p}_{\,x}>0,\\
\hat{g}_< \qquad  \hat{p}_{\,x}<0
\end{split}\right. .$$
Moreover,
$$
 \hat{\Delta} = \begin{pmatrix}
   0& \Delta (x,\,{\bf p})\\
    -\Delta^*(x,\,{\bf p}) &0
  \end{pmatrix},
  \;\hat{p}_{x}=
  \begin{pmatrix}
   p_{x,\,\alpha }& 0 \\
   0 & p_{x,\,{-\alpha}}
  \end{pmatrix},$$
where $\Delta (x,\,{\bf p})$  is the order parameter,\, and $
p_{x,\,\alpha }$ is the projection of the momentum  on the Fermi
surface on the axis $x$. Matrices  \,$\hat{v}$ have the same
structure as \,\,$ \hat{p}_{\,x}$ .

BCs for the specular reflection of electrons from the boundary:
 $ p_\parallel $ = $p_{\downarrow}\sin\vartheta_\downarrow
$ =
 $p_{\uparrow}\sin\vartheta_\uparrow =
p_S\sin\vartheta_S $,   have the form  \cite{Vod}:
\begin{gather}
(\hat{\tilde{g}}_a^S)_d = (\hat{\tilde{g}}_a^F)_d,\;\;
(\hat{\tilde{\Upsilon}}_a^S)_d =
(\hat{\tilde{\Upsilon}}_a^F)_d,\nonumber
\\
(\sqrt{\hat{R}_\alpha}-\sqrt{\hat{R}_{-\alpha}})(\hat{\tilde{\Upsilon}}_a^+)_n\nonumber
= \alpha_3(\hat{\tilde{g}}_a^-)_n,
\\
(\sqrt{\hat{R}_\alpha}-\sqrt{\hat{R}_{-\alpha}})(\hat{\tilde{\Upsilon}}_a^-)_n
= \alpha_4(\hat{\tilde{g}}_a^+)_n,\label{eq:3}
\\
-\hat{\tilde{\Upsilon}}_s^- =
\sqrt{\hat{R}_\alpha}(\hat{\tilde{g}}_s^+)_d+\alpha_1(\hat{\tilde{g}}_s^+)_n,\nonumber
\\
-\hat{\tilde{\Upsilon}}_s^+ =
(\hat{R}_\alpha)^{-\frac{1}{2}}(\hat{\tilde{g}}_s^-)_d+\alpha_2(\hat{\tilde{g}}_s^-)_n,\nonumber
\end{gather}
where\; $\hat{\tilde{g}}_{a(s)}^{\pm}
 = 1/2\,[\,{\hat{\tilde{g}}}_{a(s)}^S \pm {\hat{\tilde{g}}}_{a(s)}^F\,]$. \;
Functions $\hat{\tilde{\Upsilon}}_{a(s)}^{\pm}$ are determined
analogously. In Eq.  (\ref{eq:3}) and below the index $ d $
denotes the diagonal and $ n $ the nondiagonal part of the matrix
$( \hat{T}_{d(n)} = 1/2\,[ \, \hat{T} \pm \tau_z\hat{T}\tau_z
\,])$. GFs \,\,$\hat{\tilde{g}}$ \, are connected with GFs \,
being solutions of Eq. (\ref{eq:2}) by the following relationships
\cite{Vod}:
\begin{gather}\label{eq:4}
(\hat{\tilde{g}}_s^S)_n\,=\,(\hat{g}_s^S)_n\,\,
\cos(\,\theta_{\,\alpha}) +i\hat{\tau}_z \,(\hat{g}_a^S)_n
\,\,\sin(\,\theta_{\,\alpha})\nonumber
\\
(\hat{\tilde{g}}_a^S)_n\,=\,(\hat{g}_a^S)_n\,\,\cos(\,\theta_{\,\alpha})
+i\hat{\tau}_z \,(\hat{g}_s^S)_n\,\, \sin(\,\theta_{\,\alpha})\nonumber\\
(\hat{\tilde{g}}_s^F)_n\,=\,(\hat{g}_s^F)_n\,\,\cos(\beta_{\,\alpha}^{\,r})
+i\hat{\tau}_z (\hat{g}_a^F)_n\,\, \sin(\beta_{\,\alpha}^{\,r}) \label{eq:4}\\
(\hat{\tilde{g}}_a^F)_n\,=\,(\hat{g}_a^F)_n\,\,
\cos(\beta_{\,\alpha}^{\,r}) +i\hat{\tau}_z (\hat{g}_s^F)_n\,\,
\sin(\beta_{\,\alpha}^{\,r})\nonumber
\\
(\hat{\tilde{\Upsilon}}^{F})_n\,=\,(\hat{\Upsilon}^{F})_n\,e^{\,i\,{\rm
{sign}}\,(\hat{p}_{\,x})(\theta_{\alpha}^{\,r}+
\theta_{\,-\,\alpha}^{\,r})/2}\nonumber
\\
\theta_{\,\alpha}=\frac{\theta_{\alpha}^{\,r}-
\theta_{-\alpha}^{\,r}}{2}-
(\theta_{\,\alpha}^{\,d}-\theta_{-\,\alpha}^{\,d});\quad \,\,
\beta_{\,\alpha}^{\,r}=\frac{\theta_{\alpha}^{\,r}-
\theta_{-\alpha}^{\,r}}{2}.\nonumber
\end{gather}
The diagonal parts of matrices $\hat{\tilde{g}}$ are equal to the
corresponding matrices $\hat{g}$. The explicit form of other
functions  $\hat {\tilde{\Upsilon}}$ is not needed. These
functions are found from BCs. Coefficients $ \alpha_i$ are:
$$
\alpha_{1(2)} = \frac{1+\sqrt{{R_{\,\uparrow}}R_{\,\downarrow}}
\mp \sqrt{{D_{\,\uparrow}}D_{\,\downarrow}}}{\sqrt{R_{\,\uparrow}}
+ \sqrt{R_{\,\downarrow}}},\;\;
$$
$$
\alpha_{3(4)} = 1-\sqrt{{R_{\,\uparrow}}R_{\,\downarrow}}\, \pm
\sqrt{{D_{\,\uparrow}}D_{\,\downarrow}}\;).\;
$$
When solving Eqs. (\ref{eq:2}), let us assume that the order
parameter does not depend on the coordinate. Then for each metal
the solutions are as follows:
\begin{equation}\label{eq:5}
\hat{g}(x,{\bf p})= e^{\,-\,{\rm
{sign}}(\hat{p}_{\,x})\hat{K}x}\hat{C}({\bf p})e^{\,{\rm
{sign}}(\hat{p}_{\,x})\hat{K}x} + \hat{C}_{\,0}({\bf p})\qquad
\end{equation}
$$\hat{\Upsilon}(x,{\bf p}) = e^{\,-\,{\rm
{sign}}(\hat{p}_{\,x})\hat{K}x}\hat{\Upsilon}e^{\,-{\rm
{sign}}(\hat{p}_{\,x})\hat{K}x}, \hat{\Upsilon}=
\hat{\Upsilon}(\,x=0,{\bf p}).\,\,$$

Matrices $ \hat{C}_{\,0}({\bf p}) $ are values of  GFs $ \hat{g} $
far from the F/S boundary:
$$
\hat{C}_{\,0}^{\,S}({\bf p})\,=\,\begin{pmatrix}
  g & f \\
 f^{\,+}& -\,g
\end{pmatrix} = \frac{\begin{pmatrix}
  \varepsilon_n & -i\Delta ({\bf p}) \\
  i\Delta^*({\bf p}) & -\varepsilon_n
\end{pmatrix}}{\sqrt{\varepsilon_n^2
+ |\Delta({\bf p})|^2}}\,,
$$
\begin{equation}\label{eq:6}
\hat{C}_{\,0}^{\,F} = \hat{\tau}_z
\,\frac{\varepsilon_n}{|\varepsilon_n|},\,\, \Delta({\bf
p})\,=\,\Delta_{\,d}(T)\,\cos(2\,\vartheta_{\,S}\,-\,2\,\gamma).
\end{equation}

In Eq. (\ref{eq:6})\,\, $\vartheta_{\,S}$\, is the angle between
the electron momentum in the superconductor and the x axis
perpendicular to the contact plane and  $ \gamma $ is the angle
between the crystal \,\,$a$ \,\,axis of the  $d$-wave
superconductor and the x axis.

GFs $ \hat{g}^{\,F} $ in (\ref{eq:5}) has to tend to
$\hat{C}_{\,0}^{\,F} $ at $ x \rightarrow -\,\,\infty $ and GFs
$\hat{g}^{\,S}$ to  $ \hat{C}_{\,0}^{\,S}({\bf p})$ at $
x\rightarrow \infty $. By matrix multiplication in  (\ref{eq:5})
we find that for the above to hold it is necessary that at $x=0$
the relationship
\begin{multline}\label{eq:7}
 \hat{C}_{\,0}^{\,F}({\bf p}) \hat{C}^{\,F}({\bf p})\,=
 -\hat{C}^{\,F}({\bf p})\,\hat{C}_{\,0}^{\,F}({\bf p})=
 \,-\, {\rm {sign}}(\hat{p}_{\,x}) \hat{C}^{\,F}({\bf p})\qquad\\
\hat{C}_{\,0}^{\,S}({\bf p}) \hat{C}^{\,S}({\bf
p})\,=\hat{C}^{\,S}({\bf p})\,\hat{C}_{\,0}^{\,S}({\bf p})={\rm
{sign}}(\hat{p}_{\,x}) \,\hat{C}^{\,S}({\bf p})\quad
\end{multline}
holds. It follows from these relationships that at $x=0$:
\begin{multline}\label{eq:8}
\hat{g}_{\,s}^{\,S}\, =
\,\hat{X}\,\hat{C}_{\,a}^{\,S}\,+\,\hat{X}, \qquad
\hat{g}_{\,s}^{\,F}\, = \,\hat{C}_{\,0}^{\,F}\,-\,
\hat{C}_{\,0}^{\,F}\,\hat{C}_{\,a}^{\,F}
\\\hat{g}_{\,a}^{\,F}\, \equiv \hat{C}_{\,a}^{\,F},\qquad
\hat{g}_{\,a}^{\,S}\,=\,\hat{C}_{\,a}^{\,S}\,+\hat{C}_{\,0,\,a}^{\,S},\,\qquad
\quad
\end{multline}
where
$$ \hat{X}=
(1+\hat{C}_{\,0,\,a}^{\,S})(\hat{C}_{\,0,\,s}^{\,S})^{-\,1},\qquad
\hat{X}=\hat{\tau}_z\,(X)_d+(\hat{X})_n.
$$
In Eq. (\ref{eq:8}) $\hat{C}_{\,0,\,s(a)}^{\,S}$\,--\, are
symmetric and antisymmetric combinations of  the matrix
$\hat{C}_{\,0}^{\,S}({\bf p})$ with respect to the projection of
the Fermi momentum onto the x axis:\,\,\,\,
$\hat{C}_{\,0,\,s(a)}^{\,S}\,=\,1/2\,[\hat{C}_{\,0}^{\,S}(
p_{\,x})\,\pm \,\hat{C}_{\,0}^{\,S}(-\, p_{\,x})]$,\,
\,$\hat{X}\,=\,\hat{1}$. \,\, Matrices $\hat{\Upsilon}^{S(F)}$
satisfy the relationships:
\begin{multline}\label{eq:9}
\hat{C}_{\,0}^{\,F}({\bf p}) \hat{\Upsilon}^{F}({\bf p}) =
\hat{\Upsilon}^{F}({\bf p}) \hat{C}_{\,0}^{\,F}({\bf p}) =
\,-\,{\rm{sign}}(\hat{p}_{\,x})({\bf p}) \hat{\Upsilon}^{F}({\bf p})\\
\hat{C}_{\,0}^{\,S}({\bf p})\hat{\Upsilon}^{S}({\bf p})=
\hat{\Upsilon}^{S}({\bf p})\hat{C}_{\,0}^{\,S}({\bf p})= {\rm
{sign}}(\hat{p}_{\,x})({\bf p})\hat{\Upsilon}^{S}({\bf p}),
\end{multline}
being the condition for the functions $\hat{\Upsilon}^{F}(x,{\bf
p})$ and  $\hat{\Upsilon}^{S}(x,{\bf p})$ to tend to zero when
$x$ tends to  $-\,\infty$ and $+\,\infty$ respectively. It follows
from Eq. (\ref{eq:2}) that the function $(\hat{\Upsilon}(x)^{F})_n
= const\,=\,0$, because for a ferromagnet
$[\hat{K},(\hat{\Upsilon})_n]_+\,=\,0$ and
$\hat{\Upsilon}^{F}(x,{\bf p})$ has to tend to zero when $x$ tends
to $-\,\infty$. Then from the BCs (\ref{eq:3}) and relationships
(\ref{eq:4}) it follows that
\begin{equation}\label{eq:10}
 \alpha_3(\hat{\tilde{g}}_a^-)_n = \alpha_4
 (\hat{\tilde{g}}_a^+)_n,\qquad
\alpha_1 (\hat{\tilde{g}}_s^+)_n = \alpha_2
(\hat{\tilde{g}}_s^-)_n,
\end{equation}
From the first equality in  (\ref{eq:10}) we find the connection
between functions $(\hat{\tilde{g}}_{\,a}^{\,F})_n$ and
$(\hat{\tilde{g}}_{\,a}^{\,S})_n$:
$$(\hat{\tilde{g}}_{\,a}^{\,F})_n\,=\,
\frac{\sqrt{D_{\uparrow}D_{\downarrow}}}
{1-\sqrt{{R_{\,\uparrow}}R_{\,\downarrow}}}\,(\hat{\tilde{g}}_{\,a}^{\,S})_n.$$
By substituting this relation into the second quality in Eq.
(\ref{eq:10}) and using the relations (\ref{eq:4}) and
(\ref{eq:8}) we find $(\hat{\tilde{g}}_{\,a}^{\,F})_n$:
\begin{gather}
\hat{\tilde{g}}_a^F\,=\,\hat{g}_a^F\,e^{\,-i\,\beta_{\,\alpha}^{\,r}\,{\rm
{sign}}(\varepsilon_n)}=
-\,\frac{\sqrt{D_{\uparrow}D_{\downarrow}}\,\hat{\tau}_z\,(\hat{X})_n\,
  }{Z} \nonumber\\
Z=(1-\sqrt{R_\uparrow R_\downarrow})\,[X_d  \,
\cos(\theta_{\,\alpha})+i\, \sin(\theta_{\,\alpha})] \label{eq:11}\\
+\,(1+\sqrt{R_\uparrow
R_\downarrow})\,{\rm{sign}}(\varepsilon_n)\,
[\cos(\theta_{\,\alpha})\,+i\,X_d \,
\sin(\theta_{\,\alpha})].\nonumber
\end{gather}
From Eqs. (\ref{eq:3}) and (\ref{eq:4}) we find the rest functions
necessary to calculate conductance Eq. (\ref{eq:1}) and calculate
conductance at the ferromagnet side.

\subsection{\label{sec:level3} Differential conductance of a point FIS contact}

After carrying out the analytical continuation in these functions
(substitution $i\,\varepsilon_n $\, for\, $ \varepsilon \pm \delta
\,$ for retarded and advanced GFs, respectively), we obtain the
expression for the conductance $\sigma_{\,F/S}(V)$, which for
angles $\gamma = 0$ and $\gamma = \pi/4$ is as follows:
\begin{gather}
\sigma_{\,F/S}(V)=\frac{\,e^{\,2}\,A}{\pi}\int\frac{d{\bf
p}_{\|}\,}{\,(2\,\pi)^{\,2}}\left\{\,\,
\int\limits_{|\Delta(\vartheta_S)|}^\infty\frac{d\,\varepsilon}{2\,T}
\left[ \frac{1}{\cosh^{\,2}(\frac{\varepsilon+eV}{2\,T})}\,\right.
\right.\notag
\\
\left.  +\frac{1}{\cosh^{\,2}(\frac{\varepsilon-eV}{2\,T})}
\right]\frac{\varepsilon\,\xi^{\,R}(D_{\uparrow}+D_{\downarrow})+
\varepsilon\,(\varepsilon-\xi^{\,R})D_{\uparrow}\,D_{\downarrow}}
{Z_\Uparrow}\,\,+\label{eq:12}
\\
\int\limits_0^{|\Delta(\vartheta_S)|}\frac{d\varepsilon}{2T}
\left[
\frac{D_{\uparrow}D_{\downarrow}}{\cosh^{2}(\frac{\varepsilon+eV}{2T})}+
\frac{D_{\uparrow}D_{\downarrow}}{\cosh^{2}(\frac{\varepsilon-eV}{2T})}
\right] \left.\frac{
|\Delta(\vartheta_S)|^{2}}{Z_\Downarrow}\right\}.\notag
\end{gather}
For $\gamma = 0$:
\begin{gather}
\Delta(\vartheta_S)=|\Delta_d|\cos(2\vartheta_S)\label{eq:13}
\\
Z_\Uparrow=[\varepsilon(1-W)+ \xi
(1+W)]^{\,2}+4W\,|\Delta(\vartheta_S)|^2\sin^2(\theta_{\,\alpha})\nonumber
\\
Z_\Downarrow
=[1+2W\cos(2\theta_{\alpha})+W^{\,2}]|\Delta(\vartheta_S)|^2
-4W\varepsilon^{2}\cos(2\theta_{\alpha})\nonumber
\\
-\,\frac{16 W^2\,(|\Delta(\vartheta_S)|^2-\varepsilon^2)
\varepsilon^{2}\sin^2(2\,\theta_{\,\alpha})}
{[1+2W\,\cos(2\,\theta_{\,\alpha})+W^{\,2}]\,|\Delta(\vartheta_S)|^2
-4W\varepsilon^{2}\cos(2\theta_{\alpha})}\nonumber
\\
 W\,=\,\sqrt{R_{\uparrow}R_{\downarrow}};\qquad\xi\,=\,\sqrt{\varepsilon^2
 \,-\,|\Delta(\vartheta_S)|^2}.\nonumber
\end{gather}
For $\gamma = \pi/4$:
\begin{gather}
\Delta(\vartheta_S)=|\Delta_d|\sin(2\vartheta_S)\label{eq:14}
\\
Z_\Uparrow=[\varepsilon(1+W)+ \xi
(1-W)]^{\,2}-4W\,|\Delta(\vartheta_S)|^2\sin^2(\theta_{\,\alpha})\nonumber
\\
Z_\Downarrow
=[1-2W\cos(2\theta_{\alpha})+W^{\,2}]|\Delta(\vartheta_S)|^2
+4W\varepsilon^{2}\cos(2\theta_{\alpha})\nonumber
\\
-\,\frac{16 W^2\,(|\Delta(\vartheta_S)|^2-\varepsilon^2)
\varepsilon^{2}\sin^2(2\,\theta_{\,\alpha})}
{[1-2W\,\cos(2\,\theta_{\,\alpha})+W^{\,2}]\,|\Delta(\vartheta_S)|^2
+4W\varepsilon^{2}\cos(2\theta_{\alpha})}.\nonumber
\end{gather}
For $\gamma = 0$, при $\theta_{\,\alpha}\,=\,0$ the expression for
conductance obtained in paper \cite{Vod} follows from Eq.
(\ref{eq:12}). In the case of a nonmagnetic metal, when $
D_\uparrow = D_\downarrow $ this expression is the same as that
obtained in paper \cite{Zaitsev}, and for $D\,=\,1/(1+Z^2)$ this
expression is the same as that obtained in paper \cite{Tinkham}.
\section{Andreev reflection}
The calculation of quasiclassical GFs in the expression for
conductance allows one to conclude that for energies lower than
$|\Delta(\vartheta_S)|$ $(\varepsilon^{\,2}\,<\,|\Delta|^{\,2})$
\begin{multline}\label{eq:15}
[1-\hat{g}_{\,s}^{\,A}\, \tau_z \,\hat{g}_{\,s}^{\,R}
\,\hat{\tau}_{\,z} -\hat{g}_{\,a}^{\,A}\,\hat{\tau}_z
\,\hat{g}_{\,a}^{\,R}\,\hat{\tau}_{\,z}
 +\hat{\Upsilon}_{\,s}^{\,A}\hat{\tau}_z
\hat{\Upsilon}_{\,s}^{\,R}\,\hat{\tau}_{\,z}\\
-\hat{\Upsilon}_{\,a}^{\,A} \hat{\tau}_z
\hat{\Upsilon}_{\,a}^{\,R}
\hat{\tau}_{\,z}]=4[-\hat{\tilde{g}}_{\,a}^{\,A}\,\hat{\tau}_z
\,\hat{\tilde{g}}_{\,a}^{\,R}\,\hat{\tau}_{\,z}]\sim \hat{1}.
\end{multline}
The comparison of the form of under-gap conductances in
Eq.(\ref{eq:1}) and that of the corresponding Eq. (25) in  paper
\cite{Tinkham} shows that the matrix elements
$(\hat{\tilde{g}}_{\,a}^{\,R})^F$ and
$(\hat{\tilde{g}}_{\,a}^{\,A})^F$ are the amplitudes of the
Andreev reflection probability
$a(\varepsilon,\,\theta_{\,\alpha})$ in FIS contacts. Let us take
as $a(\varepsilon,\,\theta_{\,\alpha})$ the matrix elements of
$(\hat{\tilde{g}}_{\,a}^{\,R})^F$:
\begin{equation}\label{eq:16}
 a(\gamma,\,\varepsilon,\,\theta_{\,\alpha})= \frac{\sqrt{D_{\uparrow}D_{\downarrow}}\,\,
 \Delta(\vartheta_S)}{Z(\gamma)},
\end{equation}
 where
\begin{multline*}
Z(0)= \\(1-\sqrt{R_\uparrow R_\downarrow})[\varepsilon
\cos(\theta_{\alpha})-\sqrt{|\Delta(\vartheta_S)|^2-\varepsilon^2}
\sin(\theta_{\alpha})]
\\
+i\,(1+\sqrt{R_\uparrow R_\downarrow})
[\sqrt{|\Delta(\vartheta_S)|^2-\varepsilon^2}\cos(\theta_{\,\alpha})+
\varepsilon \, \sin(\theta_{\,\alpha})].
\\
Z(\pi/4)=\hspace{6.5 cm} \\(1+\sqrt{R_\uparrow
R_\downarrow})[\varepsilon
\cos(\theta_{\alpha})-\sqrt{|\Delta(\vartheta_S)|^2-\varepsilon^2}
\sin(\theta_{\alpha})]
\\
+i\,(1-\sqrt{R_\uparrow R_\downarrow})
[\sqrt{|\Delta(\vartheta_S)|^2-\varepsilon^2}\cos(\theta_{\,\alpha})+
\varepsilon \, \sin(\theta_{\,\alpha})].
\end{multline*}
The presence of the imaginary part in functions
$a(\gamma,\,\varepsilon,\,\theta_{\,\alpha})$  means that Andreev
reflection is accompanied by the phase shift. The Andreev
reflection probability $A_\alpha(\gamma,\varepsilon)$\quad
($A_\alpha(\gamma,\varepsilon)\,=
\,a(\gamma,\,\varepsilon,\,\theta_{\,\alpha})\,a^{\,*}(\gamma,\,\varepsilon,\,\theta_{\,\alpha})$)
is:
\begin{gather}
 A_\alpha(\gamma,\varepsilon)\,=\,
 \frac{D_{\uparrow}\,D_{\downarrow}\,|\Delta(\vartheta_S)|^{\,2}}
 {|Z(\gamma)|^2} \label{eq:17}\\
  |Z(0)|^2\,=\,[1-\sqrt{R_\uparrow R_\downarrow}]^{\,2}\,|\Delta(\vartheta_S)|^{\,2}\nonumber\\
+\,4\,\sqrt{R_\uparrow R_\downarrow}\,\,
 [\sqrt{|\Delta(\vartheta_S)|^{\,2}\,-\,\varepsilon^{\,2}}\,\cos(\,\theta_{\,\alpha})\,
+\,\varepsilon\,\,\sin(\,\theta_{\,\alpha})]^{\,2}\nonumber.\\
|Z(\pi/4)|^2\,=\,[1-\sqrt{R_\uparrow R_\downarrow}]^{\,2}\,|\Delta(\vartheta_S)|^{\,2}\nonumber\\
+\,4\,\sqrt{R_\uparrow R_\downarrow}\,\,
 [\sqrt{|\Delta(\vartheta_S)|^{\,2}\,-\,\varepsilon^{\,2}}\,\sin(\,\theta_{\,\alpha})\,
-\,\varepsilon\,\,\cos(\,\theta_{\,\alpha})]^{\,2}\nonumber.
\end{gather}
It follows from this equation that: (1) in terms of paper
\cite{Fogel} spin-mixing angle $\Theta $ for FIS contact is equal
to $\theta_{\,\alpha}$ (for SFS and NFS contacts $\Theta
$\,=\,$\theta_{\,\uparrow}^{\,r}-\theta_{\,\downarrow}^{\,r}$\,=\,
$\theta_{\,\uparrow}^{\,d}-\theta_{\,\downarrow}^{\,d}$ \quad
\cite{Barash}, \cite{Sauls2004}, \cite{Fogel}); (2) for $\gamma
=0$, for $\theta_{\,\alpha}\,<\,0$ the Andreev reflection
probability of the electron excitation with the spin projection
$\alpha$ is larger than that of the hole excitation; for
$\theta_{\,\alpha}\,>\,0$ the Andreev reflection probability of
the hole excitation with the spin projection $\alpha$ is larger
than that of the electron excitation; for $\gamma = \pi/4$ the
relation is reverse; (3) the Andreev reflection probability has
maxima at $\varepsilon\,=\,\varepsilon_{\,b}(\gamma)$ (at the
values of the energy of electron (hole) excitations corresponding
to the energy levels of Andreev surface bound states); .

The energy of bound states is:
\begin{gather}\label{eq:18}
\epsilon_{b}(0)=\left\{\begin{split}
-\,|\Delta(\vartheta_S)|\cos(\theta_{\alpha}) \,\,\, \mbox{for}\,
\,\,(\pi/2)> \theta_{\alpha}>0,
\\
|\Delta(\vartheta_S)|\cos(\theta_{\alpha})\,\,
\mbox{for}\,\,-(\pi/2)<\theta_{\alpha}<0,
\end{split}\right.
\\
\epsilon_{\,b}(\frac{\pi}{4})=\left\{\begin{split}
|\Delta(\vartheta_S)||\sin(\theta_{\alpha})|\,\,
\mbox{for}\,\,\,(\pi/2)> \theta_{\alpha}>0 \quad\\
-|\Delta(\vartheta_S)||\sin(\theta_{\alpha})| \,\, \mbox{for}
\,\,-(\pi/2)<\theta_{\alpha}<0.
\end{split}\right.\nonumber
\end{gather}
\begin{figure}[h]
\centering \epsfxsize=7.2cm \epsffile{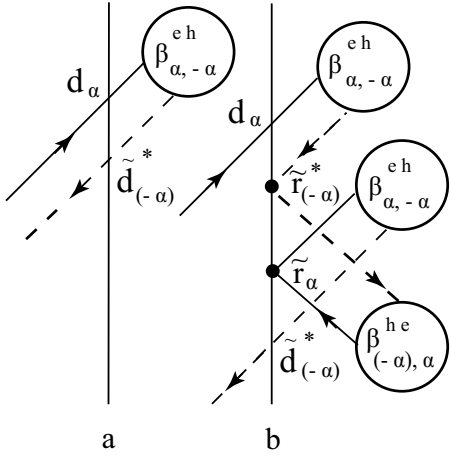} \caption{
Structure of the diagrams corresponding to Andreev reflection in
the superconductor: diagram a) one-act process; diagram b) two-act
process. The vertex $\bigcirc$ is Andreev reflection of
electron-like (solid lines) and hole-like (broken lines)
quasipaticles by the pair patential. The vertex $\bullet$ is the
normal reflection of electron-like and hole-like quasipaticles by
the barrier potential. When the solid line transforms into the
broken line,   $\bigcirc$ denotes the vertex
$\beta_{\alpha,\,-\alpha}^{\,e\,h} $. When the broken line
transforms into the solid line, $\bigcirc$ denotes the vertex $
\beta_{-\alpha,\alpha}^{\,h\,e}$. Parameters
$d_{\,\alpha},\,\tilde{d}_{\,\alpha},\,r_{\,\alpha}$ and
$\tilde{r}_{\,\alpha}$ are related as follows:
$\tilde{d}_{\,\alpha}\,=\,d_{\,\alpha}\,p_{\,x}^{\,S}/p_{\,x\,\alpha}^{\,F}$;
$\tilde{r}_{\,\alpha}\,=\,-\,r_{\,\alpha}^*\,d_{\,\alpha}/d_{\,\alpha}^*;\,\,
D_{\,\alpha}\,=\,d_{\,\alpha}\,\tilde{d}_{\,\alpha}^{\,*}$\,\,
\cite{Zaitsev}.}
\end{figure}
Andreev surface bound states are formed in a superconductor due to
the interference of electron-like and hole-like particles with
different spin-dependent phase shifts. To demonstrate this, let us
consider diagrams in Fig. 1, corresponding to Andreev reflection
of an electron with the spin projection $\alpha$ and the energy
less than $|\Delta|$ transmitted from a ferromagnet into a
superconductor. The amplitude $a(\varepsilon,\,\theta_{\,\alpha})$
is:
\begin{gather}
 a(\varepsilon,\,\theta_{\,\alpha})\,=\,
 d_{\,\alpha}\,\tilde{d}_{\,-\alpha}^*\,\beta_{\alpha,-\alpha}^{\,e\,h}[1\,+\,
 \tilde{r}_{\,-\alpha}^*\,\tilde{r}_{\,\alpha}\,
 \beta_{\alpha,\,-\alpha}^{\,e\,h}\,\beta_{-\alpha,\,\alpha}^{\,h\,e}\,\nonumber\\
+(\tilde{r}_{\,-\alpha}^*\,\tilde{r}_{\,\alpha}\,
\beta_{\alpha,\,-\alpha}^{\,e\,h}\beta_{-\alpha,\,\alpha}^{\,h\,e})^2+...]=
\,\frac{d_{\,\alpha}\,\tilde{d}_{\,-\alpha}^*\,\beta_{\alpha,\,-\alpha}^{\,e\,h}}
{1-\tilde{r}_{-\alpha}^*\tilde{r}_{\alpha}\,\beta_{\alpha,-\alpha}^{\,e\,h}\beta_{-\alpha,\alpha}^{\,h\,e}}\nonumber\\
=\frac{\sqrt{D_{\alpha}D_{-\alpha}\,p_{\,x\,\alpha}^{\,F}/p_{\,x\,-\,\alpha}^{\,F}}\,\,\,
e^{\,i\,\beta_{\,\alpha}^{\,r}}
\,\,\,\beta_{\alpha,\,-\alpha}^{\,e\,h}}
{e^{\,i\,\theta_{\,\alpha}}\,-\,
e^{\,-i\,\theta_{\,\alpha}}\sqrt{R_{\alpha}R_{-\alpha}}\,
\,\beta_{\alpha,-\alpha}^{\,e\,h}\,\beta_{-\alpha,\alpha}^{\,h\,e}}.\label{eq:19}
\end{gather}
The corresponding probability of Andreev reflection is:
\begin{equation}\label{eq:20}
    A(\varepsilon,\,\theta_{\,\alpha})\,=\,\frac{D_{\alpha}D_{-\alpha}\,p_{\,x,\,\alpha}^{\,F}/p_{\,x,\,-\,\alpha}^{\,F}
\,\,\beta_{\alpha,\,-\alpha}^{\,e\,h}\,\,\beta_{\alpha,\,-\alpha}^{*\,e\,h}}
{1+R_{\alpha}R_{-\alpha}\,\,|\beta_{\alpha,\,-\alpha}^{\,e\,h}|^{\,2}\,|\beta_{-\alpha,\alpha}^{\,h\,e}|^{\,2}-
Q}
\end{equation}
$$
Q=\sqrt{R_{\alpha}R_{-\alpha}}\left[\,\cos(2\,\theta_{\alpha})
[\beta_{\alpha,\,-\alpha}^{\,e\,h}\,\beta_{-\alpha,\,\alpha}^{\,h\,e}+
\beta_{\alpha,\,-\alpha}^{*\,e\,h}\,\beta_{-\alpha,\,\alpha}^{*\,h\,e}]\right.
$$
$$
\left.+\,i\,\sin(2\,\theta_{\alpha}) [
\beta_{\alpha,\,-\alpha}^{*\,e\,h}\,\beta_{-\alpha,\,\alpha}^{*\,h\,e}-
\beta_{\alpha,\,-\alpha}^{\,e\,h}\,\beta_{-\alpha,\,\alpha}^{\,h\,e}]\right]
$$
By comparing formulas (\ref{eq:16},\,\ref{eq:17})  with formulas
(\ref{eq:19},\,\ref{eq:20}) we find following expressions for  the
vertices $\beta_{\alpha,\,-\alpha}^{\,e\,h}$ and
$\beta_{-\alpha,\alpha}^{\,h\,e}$ for $\gamma = 0$:
\begin{equation}\label{eq:21}
   \beta_{\alpha,\,-\alpha}^{\,e\,h}=\sqrt{\frac{p_{\,x,\,-\,\alpha}^{\,F}}{p_{\,x,\,\alpha}^{\,F}}}\,\,
   \frac{\varepsilon-i\,
   \sqrt{|\Delta(\vartheta_S)|^{\,2}-
   \varepsilon^{\,2}}}{|\Delta(\vartheta_S)|}\,\frac{\Delta(\vartheta_S)}{|\Delta(\vartheta_S)|}
\end{equation}
   $$
   \beta_{-\alpha,\alpha}^{\,h\,e}\,=\,\sqrt{\frac{p_{\,x,\,\,\alpha}^{\,F}}{p_{\,x,\,-\,\alpha}^{\,F}}}\,\,
   \frac{\varepsilon\,-\,i\,\sqrt{|\Delta(\vartheta_S)|^{\,2}-\varepsilon^{\,2}}}
   {|\Delta(\vartheta_S)|}\,\frac{\,\,\Delta^*(\vartheta_S)}{|\Delta(\vartheta_S)|}.
   $$
For $\gamma = \pi/4$ the expression for the vertex
$\beta_{-\alpha,\,\alpha}^{\,h\,e}$ is of the opposite sign. It
follows from formulas (\ref{eq:20}) and (\ref{eq:21}) that in the
absence of the interferential term $Q$ the probability of Andreev
reflection is a constant (independent of the energy $\varepsilon$)
quantity. The interference of electron-like and hole-like
particles reflected by the pair potential and interface results in
the formation of Andreev surface bound states. For $\gamma = 0$ at
$\theta_{\,\alpha}\,=0\,$  the maximum in the probability of
Andreev reflection is at $\varepsilon\,=\,\pm\,|\Delta_d|$ as in
\cite{Tinkham}.  At $\theta_{\,\alpha}\,=\,\pm\,\pi/2\,$ Andreev
surface bound states with the width $\Gamma$:
\begin{equation}\label{eq:22}
    \Gamma\,=\,\frac{(1-\sqrt{R{\uparrow}\,R{\downarrow}}\,)\,|\Delta(\vartheta_S)|}
{2\,\sqrt[4]{R{\uparrow}\,R{\downarrow}}}
\end{equation}
are formed at $\varepsilon\,=\,0$  on the Fermi level. For $\gamma
= \pi/4$ the spin degeneracy of the level on the Fermi surface
\cite{Hu} at  $\theta_{\,\alpha}\,\neq 0\,$ is removed.
Electron-like and hole-like quasiparticles form separate energy
levels inside the energy gap.
\section{Appearance of Andreev bound states in conductance of the  FIS contact}
We present below the results of numerical calculations of
copnductance of the FIS contact accounting for the phase shifts.
\begin{figure}[h]
\centering \epsfxsize=7.2cm \epsffile{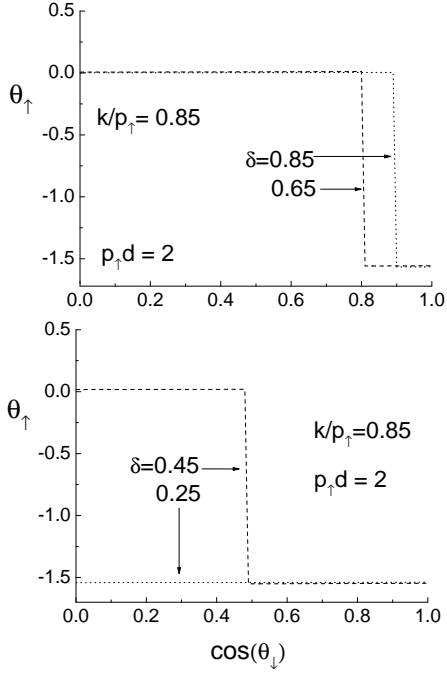}
\caption{Dependence of the angle  $\theta_{\,\uparrow}$  on
$\cos(\vartheta_{\,\downarrow})$ for various values of the
polarization of a ferromagnet $ \delta $ ($
\delta\,=\,p_\downarrow/p_\uparrow\,<1 $).}
\end{figure}
In the numerical calculations the relation between Fermi momenta
of contacting metals was the following:
$p_{\,S}\,=\,(p_{\,\uparrow}\,+\,p_{\,\downarrow})/2$.
Calculations are carried out for a rectangular barrier with the
height $U$ counted off the bottom of the conduction band of a
superconductor. The electron wave function is the isolator
$\chi(x)$ is as follows:
$$
\chi(x)\,=\,C_1\,\exp(\,\mu \,x)\,+\,C_2\,\exp(-\,\mu\,x),
$$
where
$\mu\,=\,\sqrt{k^2\,+\,p_{\,\parallel}^{\,2}};\,k^2\,=\,2m_b(U\,-\,E_F^S)\,$;
$E_F^S$   is the Fermi energy of a superconductor, $m_b$ is the
mass of an electron in a barrier.   In this case the expressions
for $ \theta_{\,\alpha}^{\,d}$ and $ \theta_{\,\alpha}^{\,r}$ have
the following form:
$$
    \theta_{\alpha}^{\,d}=\widetilde{\theta}_{\,\alpha}^{\,d}-
    i\frac{1}{2}(p_{\,x,\,\alpha}^{\,F}\,+\,p_{\,x}^{\,S})\,d;
    \quad \theta_{\,\alpha}^{\,r}\,=\,\widetilde{\theta}_{\,\alpha}^{\,r}\,-\,ip_{\,x,\,\alpha}^{\,F}\,d
$$
    \begin{equation}\label{eq:23}
    \widetilde{\theta}_{\,\alpha}^{\,d}\,=\,\arctan\left(\frac{(p_{\,x,\,\alpha}^{\,F}\,p_{\,\,x}^{\,S}-
    \mu^{\,2})\tanh(\mu \, d)}
    {\mu\,(p_{\,x,\,\alpha}^{\,F}\,+\,p_{\,x}^{\,S})} \right)
\end{equation}
    $$
    \widetilde{\theta}_{\,\alpha}^{\,r}\,=\,\arctan \left(\frac{2\,
    \mu \,
    p_{\,x,\,\alpha}^{\,F}\,[\mu^{\,2}+(p_{\,x}^{\,S})^{\,2}]
    \tanh(\mu \,d)}{Z} \right)
    $$
    $$
    Z=\mu^{\,2}\,[(p_{\,x}^{\,S})^{\,2}-(p_{x,\,\alpha}^{\,F})^{\,2}] +
    [\mu^{\,2}-p_{\,x}^{\,S}\,p_{x,\,\alpha}^{\,F}]^{\,2}\tanh^{2}(\mu
    d),
    $$
so that the angle \,\,$\theta_{\,\alpha}$\,\,\,\,
$[\theta_{\,\alpha}=(\theta_{\alpha}^{\,r}-
\theta_{-\alpha}^{\,r})/2-
(\theta_{\,\alpha}^{\,d}-\theta_{-\,\alpha}^{\,d})]\,=\,(\widetilde{\theta}_{\alpha}^{\,r}-
\widetilde{\theta}_{-\alpha}^{\,r})/2-
(\widetilde{\theta}_{\,\alpha}^{\,d}-\widetilde{\theta}_{-\,\alpha}^{\,d})$
does not depend on the location of the barrier.

Figure 2 shows the  dependences of the angle $\theta_{\,\uparrow}$
on $\cos(\,\vartheta_{\,\downarrow})$. All angles are connected by
specular reflection $ p_\parallel $ =
$p_{\downarrow}\sin\vartheta_\downarrow $ =
$p_{\uparrow}\sin\vartheta_\uparrow = p_S\sin\vartheta_S $. Figure
2 shows that the angle $\theta_{\,\uparrow},$ being a combination
of phase shifts $ \theta_{\,\alpha}^{\,d}$ and
$\theta_{\,\alpha}^{\,r},$ has a jump for a part of electron
trajectories passing through the contact region.  As the
polarization of a ferromagnet $\delta$
($\delta\,=\,p_\downarrow/p_\uparrow\,<1$) increases, the part of
electron trajectories with phase shifts experiencing a jump
increases as well and at high values of the polarization of the
ferromagnet for all electron trajectories
$\theta_{\,\uparrow}\sim\,-\pi/2,
\theta_{\,\downarrow}\sim\,+\pi/2$. For a rectangular model of the
potential barrier the angle $\theta_{\,\alpha}$ is of the order of
 $(\mp\pi/2)$  only for $k/p_{\,\uparrow}\,\leq\,1$. At $k/p_{\,\uparrow}\,>\,1$ the angle $\theta_{\,\alpha}<<1$.
\begin{figure}[h]
\centering \epsfxsize=7.2cm \epsffile{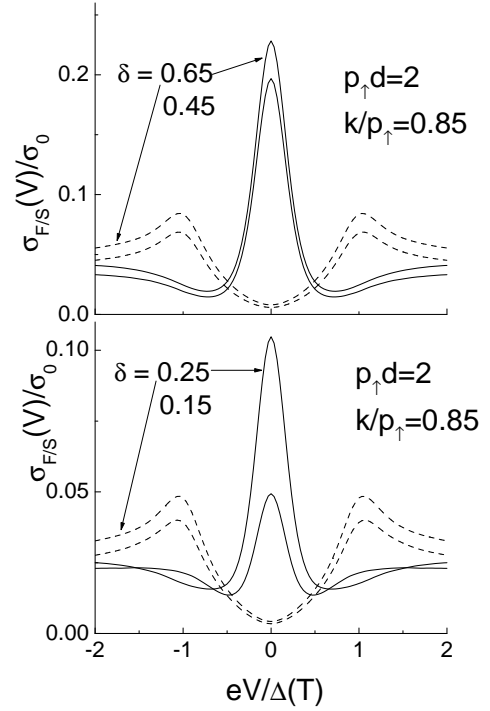}
\caption{Conductance as a function of the applied potential for (1
0 0)-oriented d - wave superconductor $(\gamma = 0)$ for various
values of the polarization of ferromagnet $\delta$ not accounting
for (dashed lines) and accounting for (solid lines) phase shifts.}
\end{figure}
\begin{figure}[h]
\centering \epsfxsize=7.2cm \epsffile{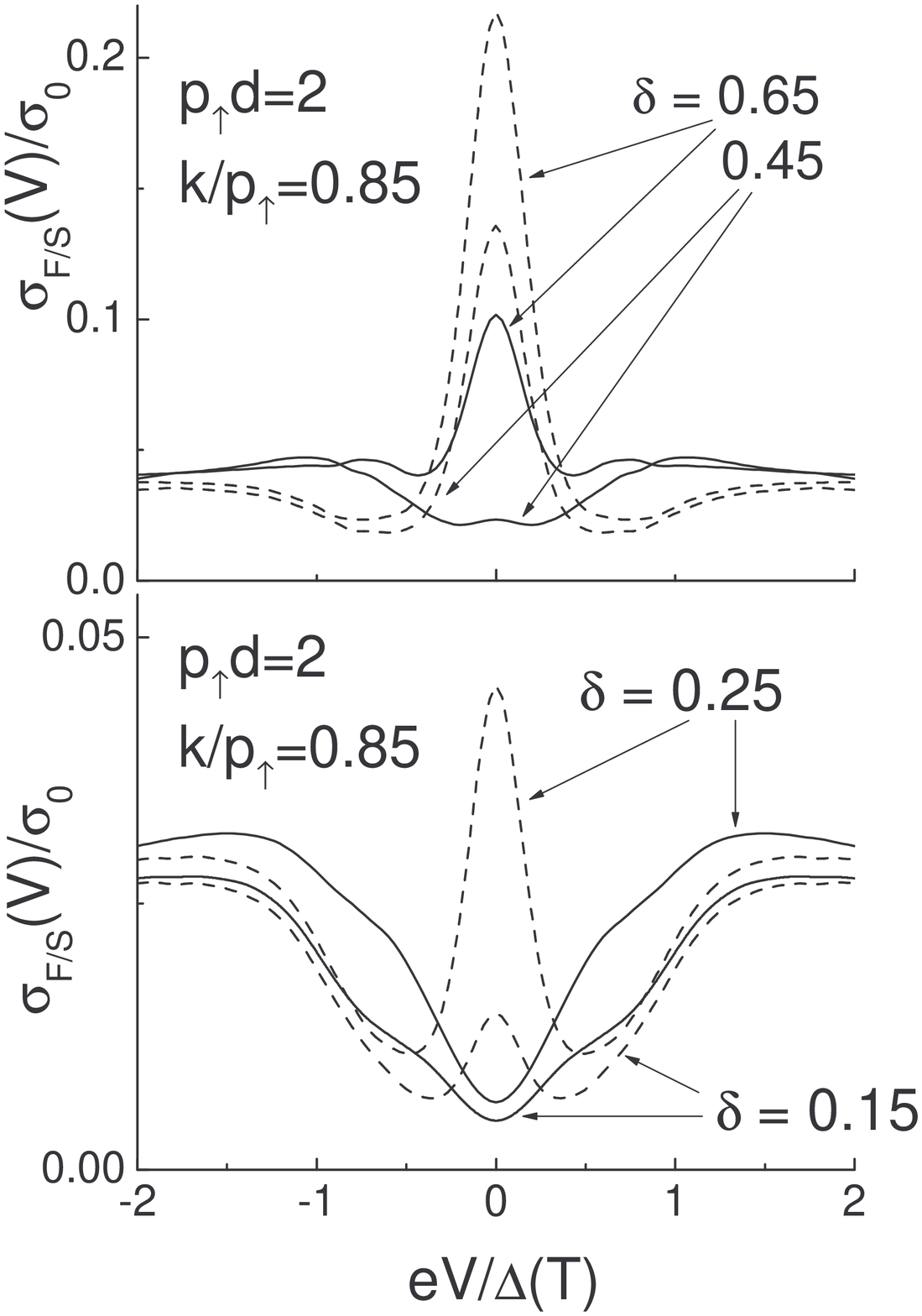}
\caption{Conductance as a function of applied potential for the (1
1 0)-oriented d - wave superconductor $(\gamma = \pi/4)$ for
various values of the  polarization of ferromagnet $\delta$ not
accounting for (dashed lines) and accounting for (solid lines) the
angle $\theta_{\,\uparrow}$. }
\end{figure}

Figure 3 shows the results of  numerical calculations of
normalized conductance of the FIS contact
$\sigma_{\,F/S}(V)/\sigma_0$ for the (100) oriented d - wave
superconductor accounting for and not accounting for phase shifts.
Not accounting for the angle ${\,\alpha}$ (dashed lines), the
plots illustrate the suppression of Andreev reflection due to the
decrease of conducting channels determined by the number of
conducting channels in the subband with a lower value of the Fermi
momentum (in this case it is $p_{F\downarrow}$) as the
polarization of ferromagnet increases. The appearance of electron
trajectories with a jump of the angle $\theta_{\,\alpha}$ form
Andreev bound states on the Fermi surface
($\varepsilon_\alpha^b(0)=0$) (\ref{eq:18}). It results in the
appearance of a peak at the zero potential in the dependence
$\sigma_{\,F/S}(V)$. As $\delta$ increases, the portion of
electron trajectories around the normal to the contact plane
participating in the formation of levels close to the Fermi level
of a superconductor increases (Fig. 1), however it does not
compensate the decrease of conductance at the zero potential due
to the decrease of conducting  channels. As a result, with
increasing polarization of ferromagnet the peak at the zero
potential in the dependence $\sigma_{\,F/S}(V)/\sigma_0$
decreases.

Figure 4 shows the results of the numerical calculations of
normalized conductance of the FIS contact
$\sigma_{\,F/S}(V)/\sigma_0$  for the (110) -oriented d - wave
superconductor $(\gamma = \pi/4)$. A part of electron trajectories
without a jump of the angle $\theta_{\,\alpha}$ forms the Andreev
bound state on the Fermi surface manifested in the conductance
peak at the zero potential. The other part of electron
trajectories with a jump of the angle $\theta_{\,\alpha}$ forms
the Andreev surface bound state  with the energy of about
$|\Delta_d\,\sin(2\vartheta_S)|$ manifested in the conductance
peak at the potential close to $|\Delta_d$. At increasing
polarization of ferromagnet all electron trajectories have a jump
of the angle $\theta_{\,\alpha}.$ As a result, the peak in
conductance at the zero potential disappears and at the potential
close to $\pm |\Delta_d\,\sin(2\vartheta_S)|$ it increases. Plots
in Fig. 4 demonstrate a tendency to the decrease of the value of
the peak in conductance at the zero potential (the decrease of the
portion of electron trajectories forming the level on the Fermi
surface) and the increase of conductance at the potential close to
the edge of the superconducting gap (the increase of the portionof
electron trajectories forming the jump of the phase shift
$\theta_{\,\alpha}$) with increasing polarization of ferromagnet.

Thus, in this paper the ballistic conductance of the point FIS
contact is calculated. The dependence of the Andreev bound states
on the spin-dependent phase shifts of electron waves reflected
from and passed through the potential barrier is found.

The influence of spin-dependent phases of electron waves passing
through and reflected by the non-magnetic potential barrier on the
Andreev reflection in the structure ferromagnet/isolator/singlet
superconductor with the order parameter having the $ d_{x^2 - y^2}
$  symmetry is studied. It is found that due to the interference
of electron-like and hole-like quasiparticles reflected by the
pair potential and interface, in a superconductor, spin-dependent
energy levels inside the superconducting gap are formed at a
distance of the order of the coherence length. The appearance of
these levels in conductance of the FIS contact is considered in
the rectangular model of the potential barrier. Features of the
dependence of the conductance of the FIS contact on the applied
potential may be used to determine the polarization of
ferromagnets by comparing these analytical expressions for the
conductance of the FIS contact with  experimental data.

The work is supported by the  Russian Foundation for Basic
Research, grant № 06-02-17233.

\end{document}